\documentclass[twocolumn,english,prl, showpacs]{revtex4}
\usepackage[T1]{fontenc}
\usepackage[latin9]{inputenc}
\usepackage{amstext}

\makeatletter
\@ifundefined{textcolor}{}
{%
 \definecolor{BLACK}{gray}{0}
 \definecolor{WHITE}{gray}{1}
 \definecolor{RED}{rgb}{1,0,0}
 \definecolor{GREEN}{rgb}{0,1,0}
 \definecolor{BLUE}{rgb}{0,0,1}
 \definecolor{CYAN}{cmyk}{1,0,0,0}
 \definecolor{MAGENTA}{cmyk}{0,1,0,0}
 \definecolor{YELLOW}{cmyk}{0,0,1,0}
 }

\makeatother

\usepackage{babel}

\begin{document}

\title{Thermodynamics of quantum measurements}

\author{Noam Erez}

\affiliation{Raymond and Beverly Sackler School of Physics and Astronomy, Tel-Aviv
University, Tel-Aviv 69978, Israel}

\date{\today}
\begin{abstract}
Quantum measurement of a system can change its mean energy, as well
as entropy. A selective measurement (classical or quantum) can be
used as a {}``Maxwell's demon'' to power a single-temperature heat
engine, by decreasing the entropy. Quantum mechanically, so can a
\emph{non-selective} measurement, despite \emph{increasing} the entropy
of a thermal state. The maximal amount of work extractable following
the measurement is given by the change in free energy: $W_{max}^{(non-)sel.}=\Delta E_{meas}-T_{Bath}\Delta S_{meas}^{(non-)sel.}$.
This follows from the {}``generalized 2nd law for nonequilibrium
initial state'' {[}Hasegawa et. al, \emph{PLA} (2010){]}, of which
an elementary reduction to the standard law is given here. It is shown
that $W_{max}^{sel.}-W_{max}^{non-sel.}$ equals the work required
to reset the memory of the measuring device, and that no such resetting
is needed in the non-selective case. Consequently, a single-bath engine
powered by either kind of measurement works at a net loss of $T_{Bath}\Delta S_{meas}^{non-sel}$
per cycle. By replacing the measurement by a reversible {}``premeasurement''
and allowing a work source to couple to the system \emph{and memory},
the cycle can be made completely reversible. 
\end{abstract}

\pacs{03.65.Ta, 03.67.-a, 05.30.-d, 05.70.Ln}

\maketitle
The thermodynamics of (selective) measurements links information theory
and physics: information on the micro-state of a system can be used
to extract more work than would otherwise be possible, and the overall
cost of acquiring, storing, deleting and using it needs to be included
in a general statement of the second law\cite{leff2003maxwell,szilard1929entropieverminderung,landauer1961irreversibility,bennett1982thermodynamics,PhysRevLett.100.080403}.
In quantum mechanics, a non-trivial measurement necessarily disturbs
the system, and can in general involve energy exchange between the
measuring device and the measured system. Therefore, in stark contrast
with the classical situation, even a generic ideal \emph{non-selective}
measurement, i.e., a measurement the outcome of which is not available
(e.g., a measurement performed by a third party, such as the environment)
will still have thermodynamic consequences. Whereas the work enabled
by selective measurements is a consequences of the increase in information
on the state (decrease of Von-Neumann entropy), a non-selective projective
measurement is always entropy non-decreasing, and unless it leaves
the state unchanged, strictly increasing\cite{breuer2002theory}.
Since a Gibbs state is the unique maximal Von-Neumann entropy state
for a given mean energy\cite{von-mathematische}, this also implies
that the such a measurement performed on a system initially in thermal
equilibrium must also increase its mean energy. Thus, a non-selective
measurement of an observable which does not commute with the thermal
state (i.e., one which does not commute with the Hamiltonian) will
take the system out of thermal equilibrium with a bath, and into a
higher-energy non-equilibrium state. This has been studied in great
detail in the case where the system itself consists of a smaller {}``system''
and a non-Markovian bath, where it was shown that even a purely dephasing
non-selective measurement that affects only the {}``system''-bath
correlation has thermodynamic consequences\cite{schulman2006ratcheting,erez2008thermodynamic}
. Such a state can be harnessed to perform work on its way back to
equilibrium with the bath. Can we determine how much work? Thermodynamics
tells us the maximal extractable work between two given states is
that performed in a thermodynamically reversible process (i.e., a
process without production of entropy, but possibly exchange of it).
Were the post-measurement state of the system, $\rho'$, a thermal
state (for some temperature), we could use textbook processes to {}``close
the cycle'', but it is not. Nonetheless, for any temperature, there
exists a Hamiltonian with respect to which $\rho'$ is a Gibbs state,
as is easy to check%
\footnote{For any inverse temperature, $\beta=\left(kT\right)^{-1}$, and density
operator, $\rho$, there is a Hamiltonian (i.e., a Hermitian operator
with spectrum bound from below) such that $\rho$ is the corresponding
Gibbs state: $\rho=Z^{-1}e^{-\beta H}$ (with $Z$ being the normalization
to trace one). For positive temperatures (finite $\beta$) this $H$
is unique up to an arbitrary additive constant and can be written
explicitly in the eigenbasis of $\rho$: $H=\beta^{-1}\sum\log\left(p_{i}^{-1}\right)|i\rangle\langle i|+E_{0}$
(where $\rho=\sum p_{i}|i\rangle\langle i|$). For zero temperature,
the restriction of $H$ to the (possibly degenerate) ground state
subspace is determined, and the rest is arbitrary, provided it commutes
with this part and its spectrum lies higher than some finite gap above
it. %
}. In particular, let $H'$ be the Hamiltonian such that $\rho'=Z^{-1}e^{-\beta H'}$.
Then performing a sudden change%
\footnote{ For the present purpose, `sudden' means much faster than the thermal
relaxation time of the state due to its coupling to the bath, and
to time scales associated with transitions due to the Hamiltonian
(actually the family of Hamiltonians the system is subject to during
the transition).%
} of the actual Hamiltonian of the system $H\mapsto H'$ brings us
to a new equilibrium state with the same bath! This non-standard (and
rapid) step is reversible, even microscopically (unitary). The equally
(macroscopically) reversible isothermal step of changing the Hamiltonian
back $H'\mapsto H$ quasi-statically, always in contact with the bath
will then return the system to the equilibrium state, closing the
cycle. The whole cycle then consists of an irreversible {}``stroke''
(the measurement) taking $\rho$ to $\rho'$ and a pair of reversible
{}``strokes'' bringing it back. The energy \emph{cost} of the measurement
is \begin{equation}
\Delta E_{meas}=\langle H\rangle_{\rho'}-\langle H\rangle_{\rho},\end{equation}
and ignorance of the outcome of the measurement (equivalently, the
fact that no sorting into subensembles corresponding to the measurement
outcomes is carried out) is responsible for the increase in the VN
entropy: \begin{equation}
\Delta S_{meas}=S\left(\rho'\right)-S\left(\rho\right).\end{equation}
In the first return stroke, the change in Hamiltonian implies the
work performed \emph{by the system} on the constraints (may be negative%
\footnote{Clearly, the addition of a constant to $H'$ leaves the corresponding
Gibbs states unchanged, so the mean value of the energy,$\langle H'\rangle_{\rho'}$,
can be made to have any desired value by appropriate choice of this
constant. In particular, it may be convenient to choose it such that
$\langle H'\rangle_{\rho'}=\langle H\rangle_{\rho'}$, making this
{}``stroke'' energy-conserving, as well as entropy-conserving. Alternately,
a choice of $\langle H'\rangle_{\rho'}=\langle H\rangle_{\rho}$ may
be useful in comparing with ref. \cite{alicki2004thermodynamics}.
In any event, we shall see that $\langle H'\rangle_{\rho'}$ will
not appear in the final result. %
}) is : \begin{equation}
W_{sudden}=\langle H\rangle_{\rho'}-\langle H'\rangle_{\rho'},\end{equation}
and the entropy is unchanged. For the last stroke we have:\begin{equation}
\Delta E_{isotherm}=\langle H'\rangle_{\rho'}-\langle H\rangle_{\rho},\end{equation}
\begin{equation}
\Delta S_{isotherm}=-\Delta S_{meas}.\end{equation}
Furthermore, the work performed during this last, isothermal, {}``stroke''
is just the change in free energy\cite{alicki2004thermodynamics}:

\begin{equation}
W_{isotherm}=\Delta E_{isotherm}-T\Delta S_{isotherm}.\end{equation}
Finally, combining these equations one gets an expression for $W_{extracted}=W_{sudden}+W_{isotherm}$
in terms of $\Delta E_{meas}$ and $T\Delta S_{meas}$:\begin{equation}
W_{extracted}=\Delta E_{meas}-T\Delta S_{meas}.\label{eq:main}\end{equation}
Which is the work extracted in a maximally reversible process (reversible
apart from the given measurement {}``stroke''), and thus the maximal
possible. More generally, this proof implies the {}``generalized
2nd law for nonequilibrium initial state''\cite{Hasegawa20101001}:

\begin{equation}
W_{ext}\geq-\Delta F_{system}\end{equation}
We have implicitly assumed that all quasi-static processes can be
performed {}``adiabatically'' (in the quantum, not thermodynamic
sense!). However, an isothermal process involving a level crossing
could introduce additional entropy production \cite{PhysRevE.71.046107}.
Such level crossings can be averted by physically exchanging the corresponding
eigenstates of the initial state $\rho$, using an nearly instantaneous
unitary right after the measurement. 

Since for a non-selective measurement, $\Delta S_{meas}>0$, as noted
above, we see that the irreversibility of the non-selective measurement
is responsible for a work deficit (compare\cite{oppenheim2002thermodynamical,alicki2004thermodynamics}
) of:\begin{equation}
\Delta W_{lost}=T\Delta S_{meas}.\label{eq:deficit_non}\end{equation}

For the selective case, let us denote the measurement basis by $\left\{ |j\rangle\right\} _{j=1,\ldots,n}$,
the corresponding values of the energy, entropy and extractable work
by: $E_{j},S_{j},W_{j}^{ext}$; and define $p_{j}=\langle j|\rho|j\rangle,\;\bar{W}=\sum p_{j}W_{j}$.
Then the maximal mean amount of extractable work will be $\bar{W}$:
\begin{eqnarray}
W_{ext}^{sel} & = & \sum p_{j}\Delta E_{j}-T\sum p_{j}\Delta S_{j}\label{eq:selmeas}\\
 & = & \Delta E-T\Delta S^{non-sel}+T\left[\Delta S^{non-sel}-\sum p_{j}\Delta S_{j}\right],\end{eqnarray}
where $\Delta S^{non-sel}=S(\rho')-S(\rho),\Delta S_{j}=S(|j\rangle\langle j|)-S(\rho)=-S(\rho)$.
The last term on the RHS of eq.\ref{eq:selmeas} is equal to the Shannon
entropy $H\left(\left\{ p_{j}\right\} \right)$, by the orthogonality
of the $|j\rangle$s \cite{nielsen2000quantum}, leading to:\begin{equation}
W_{ext}^{sel}=W_{ext}^{non-sel}+T_{bath}H\left(\left\{ p_{j}\right\} \right).\end{equation}
This is also the increase in the entropy of the auxiliary system serving
as (faithful) memory register: if the memory is initially in a pure
state, its entropy in the final state is $H\left(\left\{ p_{j}\right\} \right)$,
and for an initial pure state the entropy increase similarly to that
of the system. The minimal work required to reset the memory to its
initial state, using only the single temperature bath and a reversible
work source, is again given by the reduction in its free energy, $T_{bath}H\left(\left\{ p_{j}\right\} \right)$.
This leads to the same work deficit for a complete cycle, as in the
non-selective case:\begin{equation}
\Delta W_{lost}^{sel}=T\Delta S_{meas}^{non-sel}.\end{equation}

This begs the question whether Eq. \ref{eq:deficit_non} needs to
be corrected by deducting the work cost of resetting the memory. The
answer is negative, and follows from a remarkable property of quantum
measurements which may be called the {}``Bad apple theorem'': interaction
with a single qubit can dephase any number of qubits. More precisely,
consider a single {}``register'' qubit, labeled by the index \emph{0},
and (non-degenerate) observables $\left\{ O_{i}\right\} _{i=1,\ldots,n}$
to be measured (non-selectively) on qubits $1,\ldots,n$. If $n=1$
we can just follow the standard (and essentially only) procedure for
a \emph{selective} measurement. Namely, take the register qubit to
be in a known initial pure state, and then perform a conditional flip
to the orthogonal state (conditioned on a projection onto a particular
eigenstate of $O_{1}$, $P_{1}$). Choosing our bases such that the
initial state of the register is $|0\rangle\equiv|\sigma_{z}^{0}=-1\rangle$,
and $\left[O_{1},\sigma_{z}^{1}\right]=0$, the desired measurement
is given by the CNOT operation: \begin{equation}
U_{meas}^{(10)}=P_{+}^{1}\sigma_{x}^{0}+P_{-}^{1}I^{0},\end{equation}
where $P_{\pm}=|\sigma_{z}=\pm\rangle\langle\sigma_{z}=\pm|$ (Note
that $U^{\dagger}=U$). Viewed as a non-selective measurement, it
corresponds to a dephasing operation (corresponding to the $\sigma_{z}^{1}$
basis):\begin{equation}
\rho_{1}\mapsto\text{Tr}_{0}\left\{ U\rho_{1}\otimes\rho_{0}U\right\} =P_{+}\rho_{1}P_{+}+P_{-}\rho_{1}P_{-}.\end{equation}
The remarkable thing is, that the effect the CNOT operation has on
the state $\rho_{1}$ is exactly the same if we replace the initial
state of the register, $\rho_{0}$ by \emph{any} state that commutes
with $\sigma_{z}$ (i.e, a {}``totally dephased'' register), even
the fully mixed state! On the other hand, it is not hard to verify
that the CNOT operation transforms such states to others with this
same property, so our register is never used up! In the extreme case
where the register is totally mixed, $\rho_{0}=\frac{1}{2}I$, the
CNOT leaves the two qubits in a product state with the register qubit
unchanged! The same register qubit can therefore be used to measure
as many qubits and as often as we like. Thus, one {}``bad'' (dephased)
qubit can spoil the bunch. It is important to note that the register
does not store information corresponding to $n$ selective measurements,
so it could not be used to circumvent Bennett's argument regarding
Szilard's engine the Landauer principle. On the other hand, generically
each dephasing operation corresponds to work being performed on the
system and register, and this is the full cost of performing the measurements. 

If the measurement basis coincides with the eigenbasis of the initial
state $\rho$, $\Delta S_{meas}^{non-sel}=0$ and both cycles are
reversible (the non-selective {}``cycle'' consists of doing exactly
nothing in this case). Such a measurement corresponds to the classical
case. The reduced engine efficiency for other measurements is a quantum
feature related to the constraint that except for using the classical
result of the measurement (in the non-selective case) all operations
are restricted to the system \cite{PhysRevA.67.012320,oppenheim2002thermodynamical},
and to the fact that a measurement was implicitly defined to be irreversible.
However, the scheme described above will have the same effect, even
if we substitute a \emph{reversible} {}``premeasurement'' of the
state of the system for the measurement (i.e., a Von Neumann-type
measurement, the outcome of which is registered on a quantum detector).
Although this does not satisfy Bohr's requirement that a measurement
must include an act of irreversible amplification (\cite{bohr1958atomic},
cited in \cite{wheeler1983quantum}), the effect on the reduced state
of the system alone is identical. Such a measurement/premeasurement
is effected by a unitary operation on the system to be measured and
an auxiliary {}``quantum memory'' system, which is reversible%
\footnote{The same memory register as before, the only difference being that
no additional measurement of the state of the memory is assumed.%
}. The whole process now consists of ostensibly reversible steps provided
we include the quantum memory in our accounting. The latter does not
return to its original state, though, so the cycle is incomplete.
One may wonder whether after completing the whole scheme, one can
use the information stored in the quantum memory to recoup the work
due to its neglect (the work deficit) either directly, or by a {}``delayed
choice quantum erasure''\cite{scully1982quantum}. In fact, the neg-entropy
residing in the unread measurement outcome will be gone by the end
of the process, as far as macroscopic thermodynamics is concerned,
because the {}``reversible'' isothermal step will have entangled
the state of the memory with that of the bath. In other words, the
irreversibility of the measurement will now be attributed to the subsequent
interaction with bath. Can one circumvent this {}``quantum penalty''?
If we are allowed to jointly manipulate the system and memory register,
this is possible. Trivially, the unitary premeasurement can be undone
at short times (compared to the time of relaxation due to the bath).
More interestingly, the same effect can be achieved by an appropriate
(slowly changing) Hamiltonian for the system and memory, to protect
their correlations from thermal degradation (i.e., $H'$ now includes
an interaction term with the memory).

\begin{acknowledgments}
This work was supported by the ISF (grant no. 920/09) and the EC (FET
Open, MIDAS project). 
\end{acknowledgments}
\bibliographystyle{apsrev}
\bibliography{ThermoMeas_arxiv}

\end{document}